\documentclass[manuscript]{aastex}
\usepackage{natbib}
\usepackage{graphicx}
\def\ltsima{$\;\buildrel < \over \sim \;$}
\def\simlt{\lower.5ex \hbox{\ltsima}}
\def\gtsima{$\;\buildrel > \over \sim \;$}
\def\simgt{\lower.5ex \hbox{\gtsima}}

\newcommand{\kms}{\mbox{\,km\,s$^{-1}$}}

\newcommand\cmc{\mbox{cm$^{-2}$}}

\def \beq{\begin{equation}}
\def \eeq{\end{equation}}
\begin{document}

\title {Search for Interstellar LiH in the Milky Way}
\author{David Neufeld\altaffilmark{1}, Paul F. Goldsmith\altaffilmark{2}, Claudia Comito\altaffilmark{3} and Anika Schmiedeke\altaffilmark{3}}

\altaffiltext{1}{Department of Physics \& Astronomy, Johns Hopkins University, 3400 N. Charles St., Baltimore, MD 20218, USA}
\altaffiltext{2}{Jet Propulsion Laboratory, California Institute of Technology, 4800 Oak Grove Drive, Pasadena CA, 
91109, USA}
\altaffiltext{3}{I. Physikalisches Institut, Universit\"at zu K\"oln, Z\"ulpicher Strasse 77, D-50937 K\"oln, Germany}

\begin{abstract}

We report the results of a sensitive search for the 443.952902 GHz $J=1-0$ transition of the LiH molecule toward two interstellar clouds in the Milky Way, W49N and Sgr B2 (Main), that has been carried out using the Atacama Pathfinder Experiment (APEX) telescope.  The results obtained toward W49N place an upper limit of $1.9 \times 10^{-11}\, (3\sigma)$ on the LiH abundance, $N({\rm LiH})/N({\rm H}_2)$, in a foreground, diffuse molecular cloud along the sight-line to W49N, corresponding to 0.5\% of the solar system lithium abundance.  Those obtained toward Sgr B2 (Main) place an abundance limit $N({\rm LiH})/N({\rm H}_2) < 3.6 \times 10^{-13} \,(3\sigma)$ in the dense gas within the Sgr B2 cloud itself.  These limits 
 are considerably smaller that those implied by the tentative detection of LiH reported previously for the $z=0.685$ absorber toward B0218+357.

\end{abstract}

\section{Introduction}

Along with molecular hydrogen and its ions (H$_2^+$ and H$_3^+$) and isotopologues (e.g. HD), and together
with HeH$^+$, lithium hydride is one of a very few molecules that can be formed from the primordial elements
(H, D, He and Li) produced by big-bang nucleosynthesis.  LiH has been considered as a possible coolant of primordial material, although theoretical analyses are unanimous in concluding that the abundance of LiH in the early Universe is far too small for LiH to have played any significant role in the cooling of the gas \citep[e.g,][]{Bougleux97, Bovino11}.  

For the present Universe, containing the heavy elements that have been provided by stellar nucleosynthesis, the chemistry of LiH was first considered by \cite{Kirby78}, who identified two possible production mechanisms for LiH (and the related hydride NaH): (1) a gas-phase pathway, initiated by the slow radiative association of Li$^+$ and H$_2$ to form LiH$_2^+$ and then followed by the dissociative recombination of LiH$_2^+$ to form LiH; and (2) a grain-surface pathway, in which Li$^+$ ions are neutralized on grain surfaces and undergo reactions with hydrogen atoms on grain surfaces to form LiH.  LiH is photodissociated efficiently by the ultraviolet interstellar radiation field (ISRF). In the case of diffuse interstellar clouds, the LiH column densities predicted by \cite{Kirby78} were only $\sim 10^{6}\, \cmc$, far below the detection threshold for any conceivable spectroscopic measurement unless some significant production mechanism has been overlooked.  Larger abundances are possible in well-shielded dense clouds, particularly if the grain-surface production route is efficient.  In the case of NaH, however, \cite{Plambeck82} have obtained observational upper limits on the abundance toward several dense clouds including Sgr B2; in some sources, upper limits as small as $10^{-11}$ were obtained for the abundance of NaH relative to H$_2$, corresponding to only 10$^{-5}$ times the solar abundance of Na.   These limits led \cite{Plambeck82} to argue against grain-surface reactions being an efficient source of NaH in dense clouds.  If the same is true of LiH, then the prospects for a detection in dense clouds is extremely dim, the cosmic elemental abundance of Li being three orders of magnitude below that of Na.

Notwithstanding these pessimistic theoretical expectations, two tentative ($\sim 3 \sigma$) detections of LiH have been reported in the $z=0.685$ absorber in front of the BL Lac object B0218+357.  The first of these was obtained by \cite{Combes98}, who used the IRAM 30~m telescope to observe the $J=1-0$ transition of LiH at a redshifted frequency of 263.527~GHz, and the second by \cite{Friedel11}, who used the CARMA array to observe the same transition toward the same source.  The best-estimate LiH column densities obtained in these studies were $N({\rm LiH}) = 9 \times 10^{11}$ and $1.4 \times 10^{12}\, \cmc$ respectively.  If these tentative detections of LiH were confirmed, either in the $z=0.685$ absorber or in the Milky Way, LiH would join a relatively small list of diatomic hydrides detected in the interstellar medium.   These hydrides, comprising CH, CH$^+$, OH, OH$^+$, HF, SH, SH$^+$, HCl, HCl$^+$, and ArH$^+$, have proven invaluable diagnostic probes of the physical and chemical conditions in the interstellar gas (Gerin et al.\ 2016, and references therein).

Motivated by these possible extragalactic detections of LiH, we have performed a deep search for LiH $J = 1 - 0$ in the Milky Way galaxy.  In the local Universe (i.e.\ at $z \sim 0$) such a search requires excellent atmospheric conditions, because the LiH $J=1-0$ transition -- at a frequency of 443.952902 GHz \citep{Matsushima94} -- lies in the wing of a deep and highly pressure-broadened terrestrial water line (the $4_{23}-3_{30}$ transition at 448.001 GHz).  In this paper, we present upper limits on the 
LiH abundance thereby obtained toward 
two bright submillimeter continuum sources: Sgr B2(Main), and W49N.
Sgr B2(M) is an extremely massive and luminous molecular cloud located at a projected distance of $\sim 100$~pc on the Galactic Center, with a total mass of $\sim 2 \times 10^4\,M_\odot$ and a total bolometric luminosity of $1.2 \times 10^7\,L_\odot$ \citep{Schmiedeke16a}.  The W49N molecular cloud, located in the Perseus arm at a distance of $\sim 11$~kpc from the Sun \citep{Zhang13},
is a factor of 10 even more massive than Sgr B2(M)\citep{GalvanMadrid13}.  
Both sources exhibit a rich spectrum 
of molecular features at the systemic velocity of the background source, in addition to multiple absorption features associated with unrelated foreground material along the sight-line.  For the purposes of the present study, a search the LiH $J=1-0$ transition
in the Sgr B2(M) source provides the greatest sensitivity to LiH in dense molecular material in the Milky Way, while a search for foreground absorption in a translucent molecular cloud along the sight-line to W49N affords the best sensitivity in diffuse molecular gas.

In \S\ \ref{obs} we describe the observations and data reduction, and in \S\ \ref{results}, we present the results.  In \S\ \ref{discussion} we discuss the implications of the upper limits we have derived. 

\section{Observations and data reduction}
\label{obs}

The LiH observations were carried out using the FLASH+ receiver \citep{Klein14} and the extended bandwidth Fast Fourier Transform Spectrometer (XFFTS) backend \citep{Klein12} on the 12~m  APEX telescope \footnote{This publication is based in part on data acquired with the Atacama Pathfinder Experiment (APEX). APEX is a collaboration between the Max-Planck-Institut f\"ur Radioastronomie, the European Southern Observatory, and the Onsala Space Observatory.}\citep{Guesten06} on several nights in September 2014 under exceptionally good weather conditions.  This instrumentation yields a nominal beam size of $14^{\prime \prime}$ (FWHM) and a spectral channel width of 0.076 MHz, corresponding to 0.052 km s$^{-1}$.  Position switching observations with a clean reference position  were used to achieve the observed signal to noise ratio. For both sets of observations, the focus was checked at the beginning of each observation; pointing was checked every hour and found to be better than 2\arcsec.  The data from APEX were calibrated in units of $T^*_a$ using the standard APEX calibration tools, and smoothed to a channel width of 0.52 km s$^{-1}$ with the use of a boxcar filter.

\section{Results}
\label{results}

In Figures 1 and 2, we present the spectra obtained toward W49N and Sgr B2 (Main).  The spectra shown in the top panels have been divided by a linear fit to the continuum spectrum, and are presented on the scale of Doppler velocity relative to the Local Standard Rest (LSR) for an assumed LiH $J=1-0$ rest frequency of 443.952902 GHz \citep{Matsushima94}.  
A set of emission lines, observed toward both sources, can be identified as a series of $K_A = 5 \rightarrow 4$ rotational transitions of $\rm ^{34}SO_2$, all with $\Delta J = 0$. Blue vertical lines, labeled with the upper rotational quantum number $J$, indicate positions of these emission lines for assumed centroid velocities of +10 (W49N) and $+59\kms$ (Sgr B2 Main) relative to the LSR.  The stronger emission features appearing near +120 (W49N) and $+170 \kms$ (Sgr B2 Main) are the 443.79176~GHz $31_{3,39} - 31_{2,30}$ transition of (the dominant isotopologue) $\rm ^{32}SO_2.$  A series of $J=21 \rightarrow 20$ emission lines of ketene ($\rm H_2CCO$), marked by red vertical lines, is also evident in the spectra.

\begin{figure}
\includegraphics[width=11 cm]{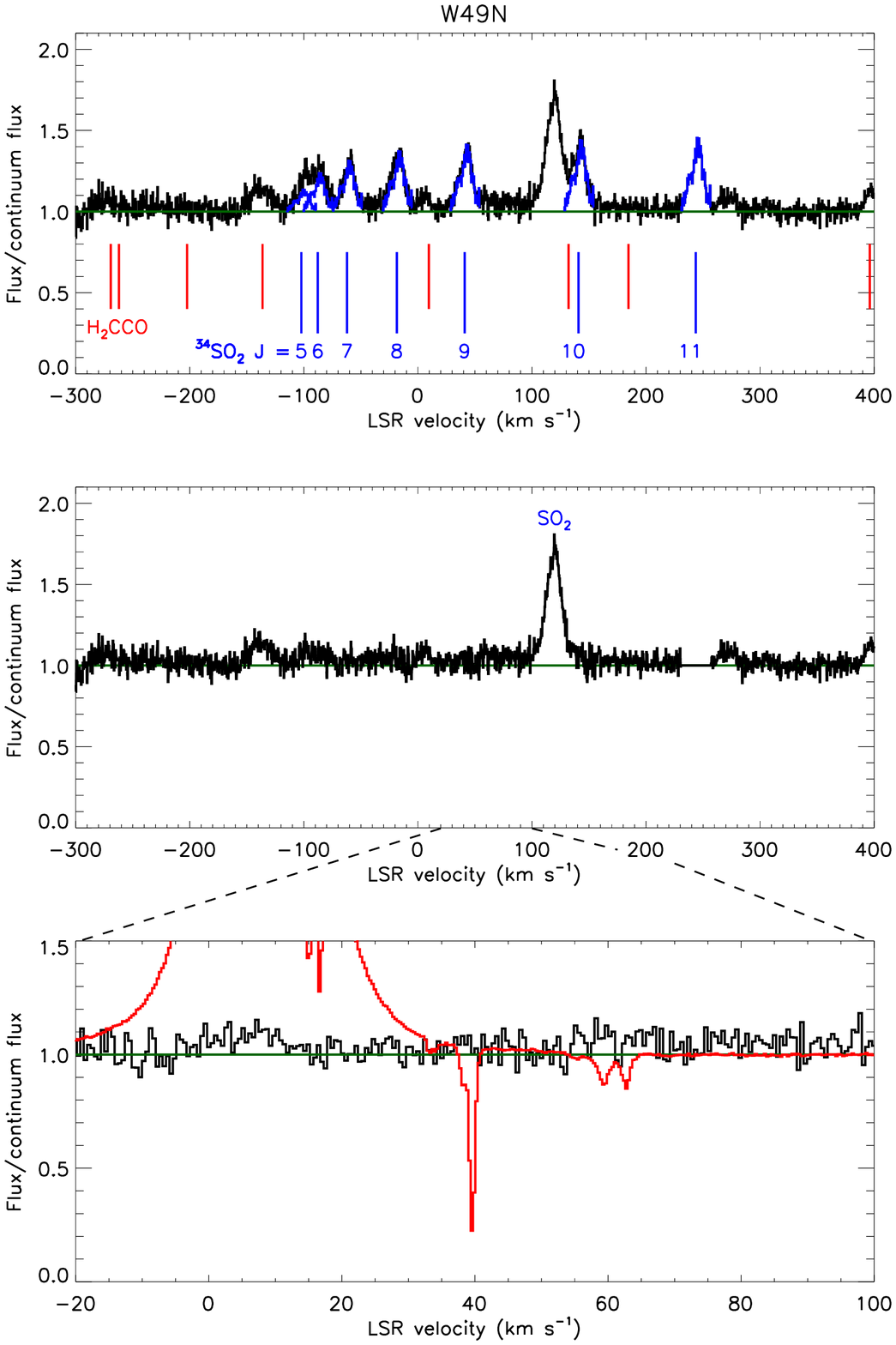}
\caption{LiH spectrum obtained for W49N.  Top panel: continuum-divided spectra, in which a set of emission lines can be identified as a series of $K_A = 5 \rightarrow 4$ rotational transitions of $\rm ^{34}SO_2$, all with $\Delta J = 0$. Blue vertical lines, labeled with the upper rotational quantum number $J$, indicate positions of these emission lines for an assumed centroid velocity of $+10\kms$ relative to the LSR, while red vertical lines indicate the positions of $\rm H_2CCO$ emission lines.
Middle panel: continuum-divided spectra after removal of $\rm ^{34}SO_2$ emission lines.  
The remaining strong emission feature near $+120 \kms$ is the 443.79176~GHz $31_{3,39} - 31_{2,30}$ transition of (the dominant isotopologue) $\rm ^{32}SO_2$.
Bottom panel: expanded view of middle panel, with
H$_2$S $1_{10}-1_{01}$ spectrum superposed in red \citep{Neufeld15}.
 }
\end{figure}

\begin{figure}
\includegraphics[width=10 cm]{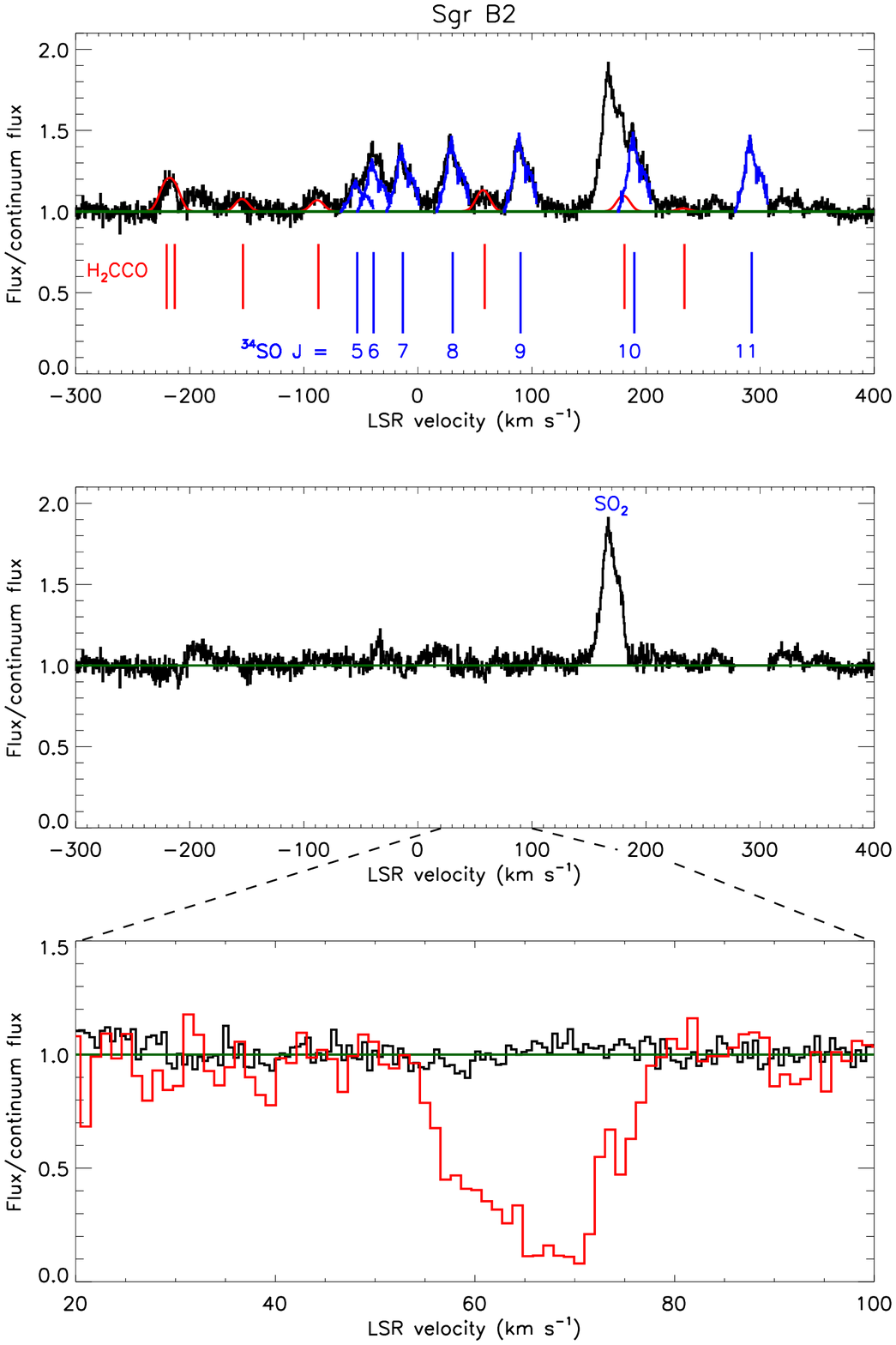}
\caption{LiH spectrum obtained for Sgr B2 (M).  Top panel: continuum-divided spectra, in which a set of emission lines can be identified as a series of $K_A = 5 \rightarrow 4$ rotational transitions of $\rm ^{34}SO_2$, all with $\Delta J = 0$. Blue vertical lines, labeled with the upper rotational quantum number $J$, indicate positions of these emission lines for an assumed centroid velocity of $+59\kms$ relative to the LSR, while red vertical lines indicate the positions of $\rm H_2CCO$ emission lines.  Middle panel: continuum-divided spectra after removal of $\rm ^{34}SO_2$, $\rm H_2CCO$ and $\rm ^{34}SO$ emission lines (see text).  
The remaining strong emission feature near $+170 \kms$ is the 443.79176~GHz $31_{3,39} - 31_{2,30}$ transition of (the dominant isotopologue) $\rm ^{32}SO_2$.
Bottom panel: expanded view of middle panel.  For comparison, the red histogram in the bottom panel shows the profile of a strong absorption feature in the bandpass, the
$\rm CH_3OH$ $3_{12}-2_{02}$ line at 445.571~GHz.
 }
\end{figure}

We have removed the $\rm ^{34}SO_2$ emission lines appearing in the upper panel of Figures 1 and 2 -- together with additional weak emission lines in the case of Sgr B2 (M) -- to obtain the spectra shown in the middle panel.  For the $\rm ^{34}SO_2$ lines, we adopted the $J_{K_A,K_C} = 11_{5,7}-11_{4,8}$ transition as a template, and assumed that a single rotational temperature could account for the $\rm ^{34}SO_2$ line ratios.  An excellent fit (blue curves in the top panel) was obtained for an assumed rotational temperature of 100~K for W49N and 60~K for Sgr B2.  In the case of Sgr B2 (M), an extensive line survey has been conducted at millimeter wavelengths by  \cite{Belloche13}, who derived estimates of the source size, column density and rotational temperature for 46 molecules and 54 minor isotopologues thereof.  Based upon these estimates, we have used the XCLASS software \citep{Moeller17} to compute a predicted spectrum for the weak ketene lines in the spectral region targeted in the present study, together with one weak line of $\rm ^{34}SO$ that also lies in the bandpass.  This predicted spectrum is shown by the red curve in the top panel of Figure 2, and was subtracted along with the $K_A = 5 \rightarrow 4$ rotational transitions of $\rm ^{34}SO_2$ to obtain the middle panel in that figure. 

The bottom panels in Figures 1 and 2 show zoomed versions of the spectra after subtraction of the $\rm ^{34}SO_2$ emission lines (and, in the case of Sgr B2 (M), of the predicted ketene and $\rm ^{34}SO$ spectrum.)   For comparison, in Figure 1, the red histogram in the bottom panel shows the H$_2$S $1_{10} - 1_{01}$ spectrum obtained by \cite{Neufeld15} for W49N.  As is observed for many molecules \citep[e.g.,][]{Neufeld15}, an absorbing cloud at $v_{\rm LSR} \sim 40\kms$ gives rise to a strong absorption line in the H$_2$S spectrum.  The LiH spectrum, by contrast, shows no hint of any absorption feature at $v_{\rm LSR} \sim 40\kms$.  There appears to be a weak emission feature, perhaps of marginal significance, close to the systemic velocity of source.  Even if real, this could not be identified with LiH $J=1-0$, because ground-state rotational lines observed at submillimeter wavelengths in this and other similar sources are expected and indeed observed to appear in absorption.  A blend of the $22_{5,18} - 21_{5,17}$ and $22_{5,17} - 21_{5,16}$ lines of ketene is a plausible identification.

\section{Discussion}
\label{discussion}

\subsection{Upper limit on LiH for a diffuse molecular cloud, derived from observations of W49N}

\begin{figure}
\includegraphics[width=15 cm]{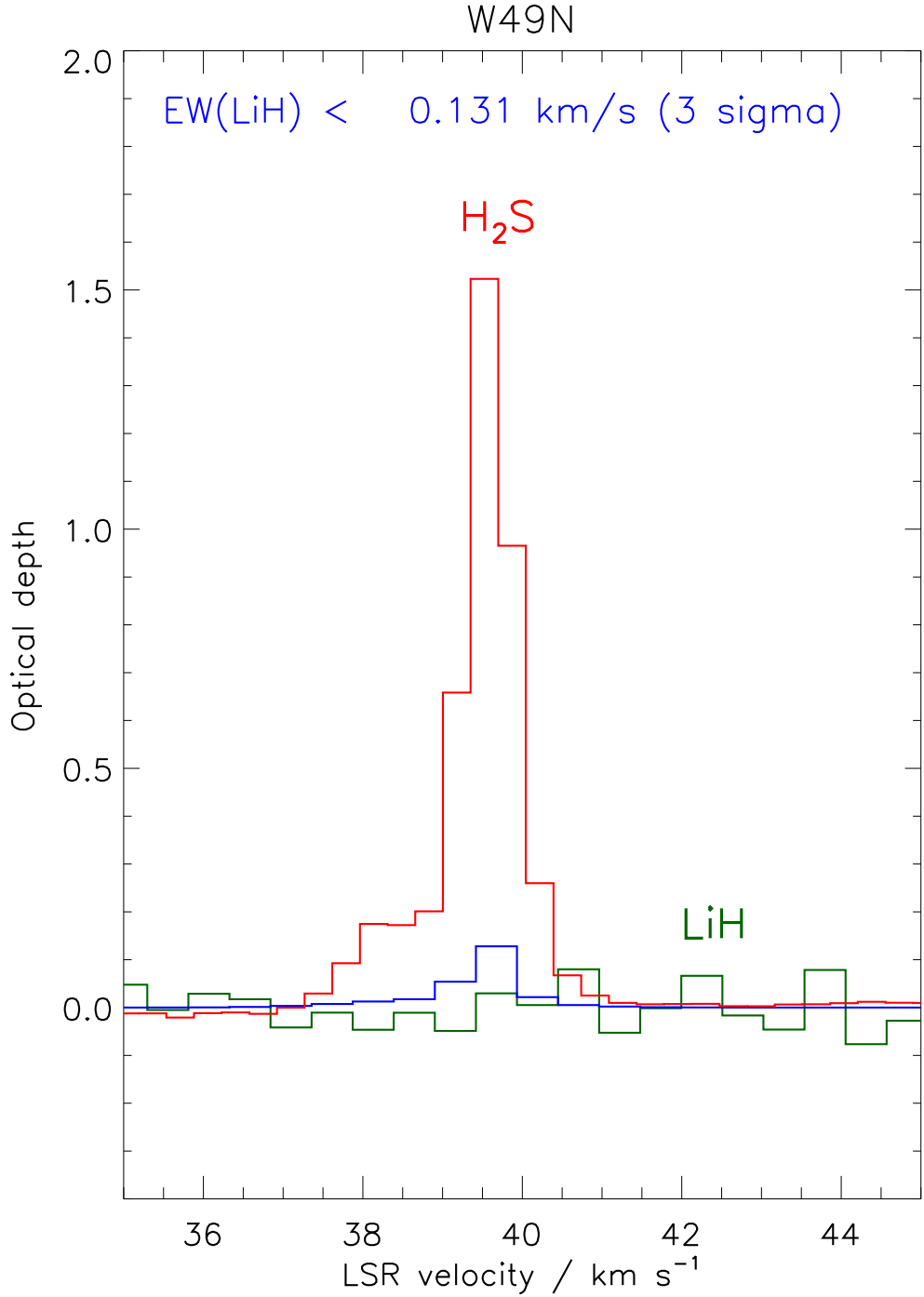}
\caption{Optical depths for H$_2$S $1_{10} - 1_{01}$ (red) and LiH $J=1-0$ (green), as a function of $v_{\rm LSR}$.  The blue histogram shows the LiH optical depth that would be obtained for an LiH abundance equal to the 3 $\sigma$ upper limit, $1.9 \times 10^{-11}$ relative to H$_2$, which yields an absorption line equivalent width (EW) of 0.131 km$\rm \,s^{-1}$.}
\end{figure}

For the W49N spectrum, we have obtained an upper limit on the LiH column density in the $v_{\rm LSR} = +40 \rm \, km \, s^{-1}$ absorbing cloud by using the H$_2$S $1_{10} - 1_{01}$ line profile as a template.  In Figure 3, we show the line optical depths for H$_2$S $1_{10} - 1_{01}$ (red) and LiH $J=1-0$ (green), as a function of $v_{\rm LSR}$. 
The spectra have been regridded onto a common velocity grid.  Assuming a LiH line profile identical to that of H$_2$S, we used a $\chi^2$ analysis to place a $3 \sigma$ upper limit on the LiH optical depth (blue).  To determine that limit, we considered the mean square deviations between the observed data and a scaled version of the H$_2$S spectrum, and adjusted the scaling factor so as to increase the value of $\chi^2$ by 9.  The resultant upper limit on the integrated optical depth of the LiH line is $ \int \tau dv \le 0.131 \, \rm km\,s^{-1}\,(3 \sigma).$  At the low densities and radiation fields typical of diffuse molecular clouds, most LiH molecules will be in the ground rotational state, and the rotational temperature will lie close to the temperature of the cosmic microwave background.  Adopting a dipole moment for LiH of 5.88~Debye \citep{Buldakov04}, which leads to a large spontaneous decay rate of $1.1 \times 10^{-2}\,\rm s^{-1}$, and assuming an excitation temperature of 2.73~K, we obtain the following upper limit on the LiH column density:
$$ N({\rm LiH}) = 2.52 \times 10^{11} \, {\rm cm}^{-2} \int \tau dv / \rm km\,s^{-1} \le 3.3 \times 10^{10} \, {\rm cm}^{-2} \,(3\sigma).$$  The H$_2$ column density in this cloud has been estimated as $1.76 \times 10^{21} \, {\rm cm}^{-2},$ so our $3\sigma$ upper limit on $N({\rm LiH})$ corresponds to a LiH abundance of $1.9 \times 10^{-11}$ relative to H$_2$.  By comparison, the elemental abundance of lithium nuclei in the solar system is $1.8 \times 10^{-9}$ relative to H {\it nuclei} \citep{Asplund09}, so our upper limit on $N({\rm LiH})/N({\rm H_2})$ is equivalent to 0.5$\%$ of the solar system lithium abundance.

This limit is perhaps unsurprising, since most hydrides -- defined here as molecules containing one heavy element atom and one or more hydrogen atoms -- are only minor reservoirs of heavy elements in diffuse molecular clouds 
\citep{Gerin16}.  For most elements, including O, C, S, and N, hydrides typically account for no more than 0.1$\%$ of the gas-phase abundance.  The notable exceptions are HF, which typically accounts for $\sim 40\%$ of the gas-phase fluorine abundance in diffuse molecular clouds, and HCl, HCl$^+$ and H$_2$Cl$^+$, which account for up to $\sim 10\%$ of the gas-phase chlorine \citep{Gerin16}.  These exceptions are readily understood by considerations of thermochemistry, in that F and Cl are the only elements that react exothermically with H$_2$ in their dominant ionization stage in diffuse clouds (F and Cl$^+$), thereby forming HF and HCl$^+$ rapidly at low temperature.  In the case of LiH, by contrast, no similarly rapid formation process has been proposed.  As discussed previously, the formation of LiH is expected to be slow, proceeding by via grain surface reactions or in a gas-phase chemical sequence initiated by the slow radiative association of Li$^+$ and H$_2$.

\begin{figure}
\includegraphics[width=14 cm]{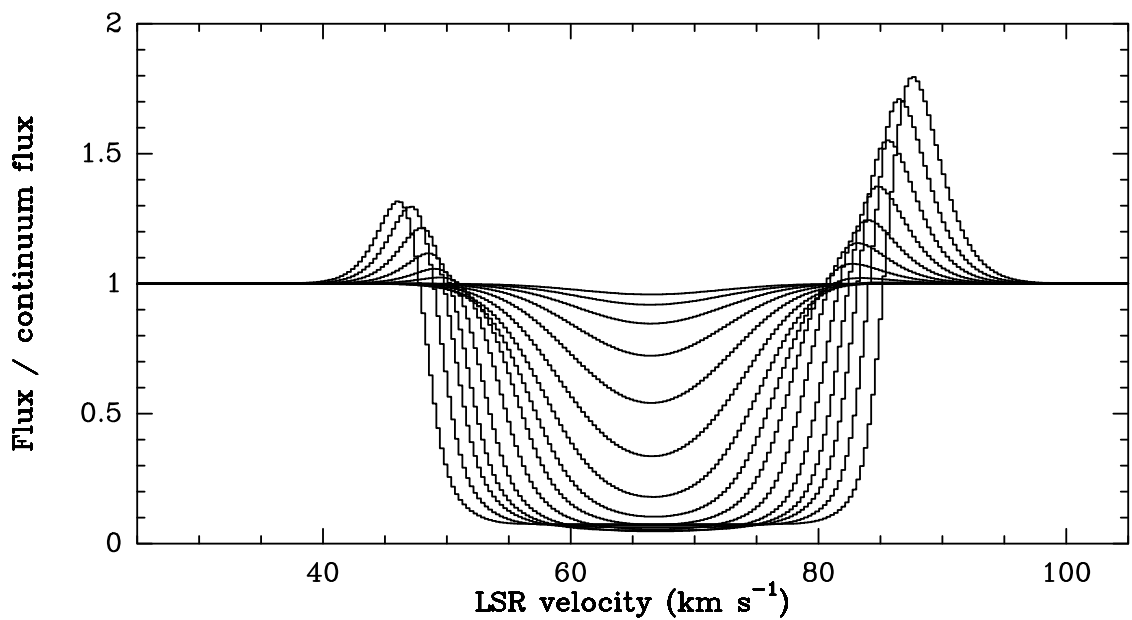}
\caption{Line profiles for Sgr B2 (Main), as predicted using the model described in the text.   Predictions are shown for $n({\rm LiH})/n({\rm H}_2)$ ratios ranging from $2.22 \times 10^{-13}$ to $3.64 \times 10^{-9}$ (the maximum value corresponding to the solar system lithium abundance), in steps of a factor 2.  (The absorption line grows deeper and then broader with increasing $n({\rm LiH})/n({\rm H}_2)$; for $n({\rm LiH})/n({\rm H}_2) \simgt 10^{-10}$, the predicted profile exhibits significant emission in the line wings.) }
\end{figure}

\begin{figure}
\includegraphics[width=15 cm]{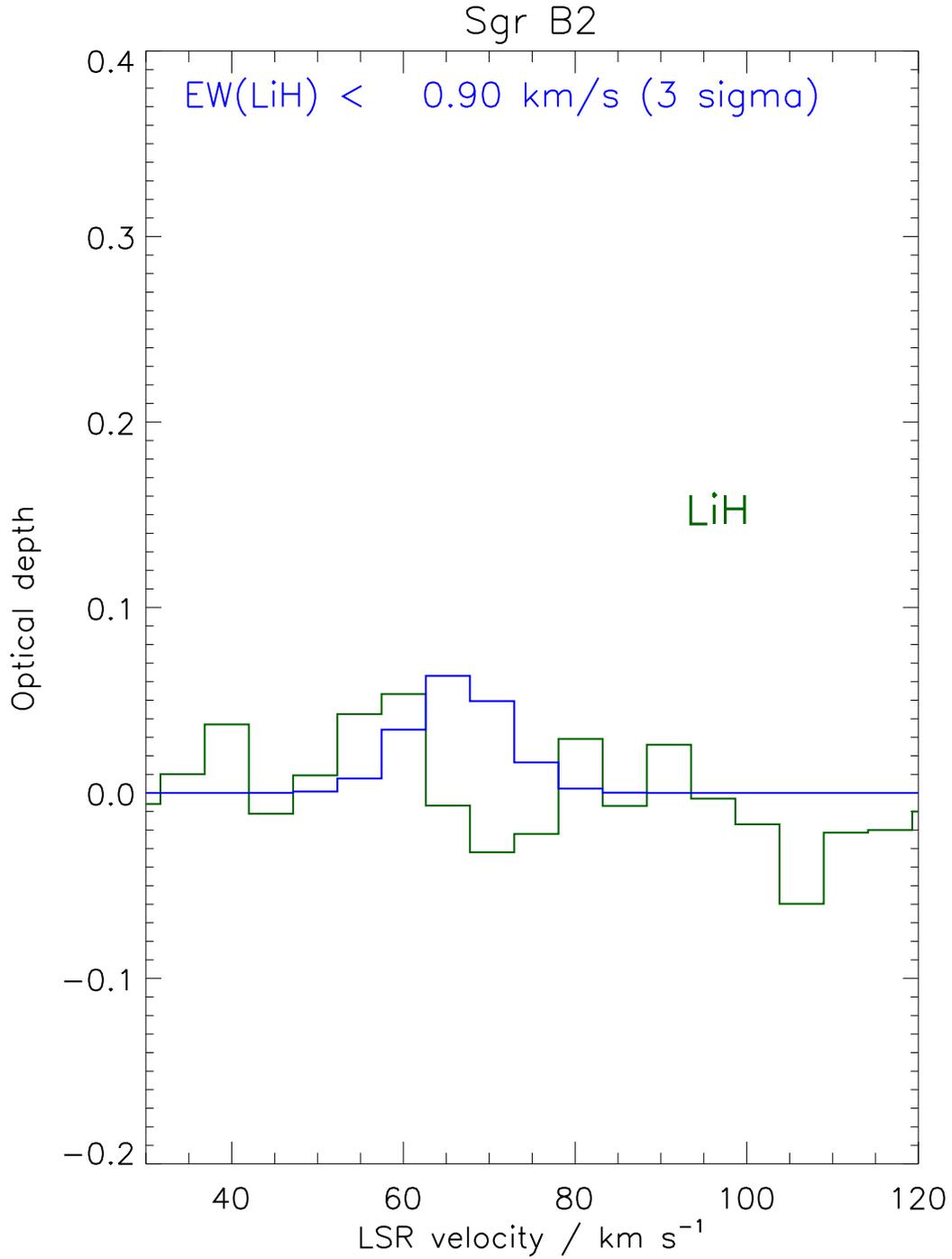}
\caption{Optical depth for LiH $J=1-0$ (green), as a function of $v_{\rm LSR}$.  The blue histogram shows the LiH optical depth that would be obtained for an LiH abundance equal to the 3 $\sigma$ upper limit, $3.6 \times 10^{-13}$ relative to H$_2$, which yields an absorption line equivalent width (EW) of 0.90 km$\rm \,s^{-1}$}
\end{figure}

\vskip 1 true in
\subsection{Upper limit on LiH for a dense molecular cloud, derived from observations of Sgr B2}

From the Sgr B2 (Main) spectrum, we may obtain a lower limit on the LiH abundance in the background source itself.  In this case, the gas densities are much higher than in the diffuse ISM, and collisional excitation cannot be neglected.  Moreover, a strong far-infrared radiation field can lead to significant radiative pumping.  A model for the excitation of LiH and the radiative transfer of its line radiation is therefore needed. 

Here, we employed the Python tool ``Pandora", developed by \cite{Schmiedeke16b}, to make coordinated use of the RADMC-3D \citep{RADMC-3D} and LIME \citep{LIME} radiative transfer codes.   Given an assumed density profile and a distribution of heating sources, 
RADMC-3D provides the temperature profiles that LIME uses as input for the line radiative transfer calculations. Our density and temperature profiles for the Sgr B2 cloud correspond to Model C of \cite{Schmiedeke16a}, which provides a good fit to the observed infrared continuum radiation. Excitation calculations were performed for the lowest 11 rotational states of LiH, with the use of  spontaneous radiative rates from \cite{Plummer84}.  Calculated rate coefficients are not available for collisional de-excitation of LiH, so we adopted rate coefficients for the de-excitation of HCl by H$_2$ \citep{Lanza14} and computed the excitation rates using the principle of detailed balance.  In the optically-thin regime of present interest, the line was predicted to be entirely in absorption, with a predicted equivalent width that was found to be proportional to the adopted abundance. We assumed a constant gas-phase abundance for LiH along the line-of-sight, thus neglecting fluctuations linked to freeze-out onto or evaporation from dust grains. A detailed description of the parameters adopted for the spectral line modeling of Sgr B2 with ``Pandora" can be found in \cite{Schmiedeke16b} and in Comito et al.\ (in preparation.)

We computed the expected line strengths and profiles as a function of the $n({\rm LiH})/n({\rm H}_2)$ abundance ratio.  In Figure 4, we show a set of predicted line profiles for several values of $n({\rm LiH})/n({\rm H}_2)$, ranging from $2.22 \times 10^{-13}$ to $3.64 \times 10^{-9}$ (the maximum value corresponding to the solar system lithium abundance), in steps of a factor 2.  We then used a $\chi^2$ analysis to determine the largest abundance consistent with the data, and thereby obtained an upper limit $n({\rm LiH})/n({\rm H_2}) < 3.6 \times 10^{-13}\, (3\sigma)$.  In Figure 5, we show the observed spectrum (green), and the expected line profiles for the upper limit given above (blue).

Our upper limit on the LiH abundance along the sight-line to Sgr B2 (Main) corresponds to a fraction $\sim {\bf 9} \times 10^{-5}$ of the solar system lithium abundance.  This fraction is similar to the fraction of the solar sodium abundance corresponding to the typical NaH limits obtained by \cite{Plambeck82}.  Thus, our upper limit on LiH suggests that Plambeck \& Erickson's conclusion that NaH cannot be efficiently be formed on grain surface applies also to LiH.  It also places constraints on any rapid gas-phase production route that may have been overlooked in previous theoretical studies.

\subsection{Comparison with the tentative detection reported toward B0218+357}

Based upon their tentative detection of LiH toward B0218+357 with $N({\rm LiH}) = 1.4 \times 10^{12} \, \cmc$ \cite{Friedel11} inferred a $N({\rm LiH})/N({\rm H_2})$ abundance ratio in the range $2.9 \times 10^{-12}$ to $2.9 \times 10^{-10}.$  The large uncertainty in the derived abundance reflects uncertainties in the column density of {\it molecular hydrogen}, for which previous literature estimates ranged from $5 \times 10^{21}$ to $5 \times 10^{23}\,\cmc$ \citep{Menten96, Combes95}.  At the bottom of that range, the absorbing cloud is most nearly similar to the diffuse molecular cloud in front of W49N, but the inferred abundance for B0218+357 is 15 times our 3$\sigma$ upper limit.  At the top of that range, the comparison with Sgr B2 is most appropriate.  Here, our 3 sigma upper limit lies a factor of $\sim 8$ below the inferred abundance for B0218+357.  In either case, the upper limits derived in the present study suggest that the chemical conditions in the Milky Way are significantly different from -- and less favorable for LiH formation than -- those in the B0218+357 absorber, or else that the previously-reported tentative detections toward the latter source are in fact spurious.  Deeper integrations toward B0218+357 would be desirable to resolve the question of whether the previous detections are real.

\begin{acknowledgements}

We  express our gratitude to Miguel Requena--Torres and the staff of the APEX telescope for their able assistance with these observations.  We thank Arnaud Belloche for providing us with the
\cite{Belloche13} model parameters in digital (XCLASS) format.  This research was conducted in part at the Jet Propulsion Laboratory, which is operated by the California 
Institute of Technology under contract with the National Aeronautics and Space Administration (NASA). 

\end{acknowledgements}

\newpage
\bibliography{bibdata}
\end{document}